\documentclass[twocolumn,showpacs,amsmath,amssymb]{revtex4}
\usepackage{graphicx}
\begin{document}
\title {\Large \bf Rotating Spacetimes with Asymptotic Non-Flat Structure and the Gyromagnetic Ratio\footnote{\normalsize To the
memory of Erdal In\"{o}n\"{u}.}}
\author{\large Alikram N. Aliev}
\address{Feza G\"ursey Institute, P. K. 6  \c Cengelk\" oy, 34684 Istanbul, Turkey}
\date{\today}

\begin{abstract}

In general relativity, the gyromagnetic ratio for all stationary, axisymmetric  and asymptotically flat  Einstein-Maxwell fields is  known to be $ g=2 $. In this paper, we continue our previous works of examination this result for  rotating charged spacetimes with asymptotic non-flat structure. We  first consider  two instructive examples of these spacetimes: The  spacetime of a Kerr-Newman black hole with a straight cosmic string on its axis of symmetry and the Kerr-Newman Taub-NUT (Newman-Unti-Tamburino) spacetime. We show that for both spacetimes the gyromagnetic ratio $ g=2 $ independent of their asymptotic structure. We also extend this result to a general class of metrics which admit separation of variables  for the Hamilton-Jacobi and wave equations. We proceed with the study of the gyromagnetic ratio in higher dimensions by considering  the general solution for rotating charged black holes in minimal five-dimensional  gauged supergravity. We obtain the analytic  expressions for two distinct gyromagnetic ratios of these black holes that are associated with their two independent rotation parameters. These expressions reveal the dependence of the gyromagnetic ratio on both the curvature radius of the AdS background and the parameters of the black holes: The mass, electric charge and two rotation parameters. We explore some special cases of interest  and show that when the two rotation parameters  are equal to each other and the rotation  occurs at the maximum angular velocity, the gyromagnetic ratio  $ g=4 $ regardless of the value of the electric charge. This  agrees precisely with  our earlier result  obtained for general Kerr-AdS black holes  with a test electric charge. We also show that in the Bogomol'nyi-Prasad-Sommerfield (BPS) limit the gyromagnetic ratio for a supersymmetric black hole with equal rotation parameters  ranges  between  2 and 4.

\end{abstract}

\pacs{04.20.Jb, 04.70.Bw, 04.50.+h}

\maketitle

\section{Introduction}

In the theory of electromagnetism, it is well known fact that rotation of a uniformly charged body creates a stationary magnetic field. The magnetic field has a dipole character and its far distance behavior
determines the {\it magnetic moment} of the body. There is a simple relation between the  magnetic dipole moment and  the total angular momentum of the body and the  constant of proportionality in this relation defines the gyromagnetic ratio $ g $ . In all cases, when the motion is slow and the charge-to-mass  ratio is constant, the gyromagnetic ratio $ g=1 $. However, this result drastically changes for a charged quantum particle. The most direct manifestation of this appears in the  non-relativistic limit of the Dirac equation in a uniform magnetic field. It turns out that for an  electron the gyromagnetic ratio  corresponds to $ g=2 $. Furthermore, it is now well justified that, up to radiative corrections, the value $ g=2 $ is the natural gyromagnetic ratio for  charged quantum particles of arbitrary spin \cite{weinberg, fmt, jakiw1, jakiw2}.

It is a striking fact that in general relativity rotating charged black holes possess the same value of the gyromagnetic ratio as an electron and other elementary particles \cite{carter1}. As is known, the black holes are uniquely characterized  by their independent physical parameters, the  mass, angular momentum and the electric charge. It is also known that in some respects  a  rotating charged black hole behaves like an ordinary electrodynamical object. One therefore may expect that the black hole must have a magnetic dipole moment determined by its independent physical parameters . The ratio of the magnetic moment to the associated  angular momentum is given by the $g$-factor of $ 2 $, unlike the case of a rotating uniformly charged  matter for which $ g=1 $. Subsequently, it has been shown  that $ g=2 $ is related to the internal symmetry of the system and it remains unchanged for all stationary, axisymmetric  and asymptotically flat solutions of the Einstein-Maxwell equations (see for instance, Refs. \cite{sim, rt, bahram}). The gyromagnetic ratio has also been examined for a loop of charged matter rotating around a static charged black hole. The analysis has shown that for large radii of the loop, the gyromagnetic ratio reduces to $ g = 1  $ , whereas it approaches the general-relativistic value $ g = 2 $   when the loop merges with the event horizon of the black hole \cite{garf}.

Recently, there has been a renewed interest in the gyromagnetic ratio of charged rotating systems in general relativity. In \cite{pfister}, the authors considered a spherical model consisting of a charged mass shell in the limit of slow rotation. It was found that in a wide range of the model parameters determined by the mass, charge, and radius of
the system, the gyromagnetic ratio remained near the $ g = 2 $. On the other hand, it has been argued in \cite{novak} that  $ g = 2 $ can not be attained in a model for rapidly rotating and charged compact stars. Moreover, it turns out that  $ g \neq 2 $ for a relativistically rotating charged disk whose electromagnetic field  in a certain limit  is described by the "magic" field of the Kerr-Newman black hole when the gravitational interaction is turned off \cite{bell}. In a more general context of spacetimes fulfilling  Carter's separability ansatz, the gyromagnetic ratio was  studied in \cite{gair}.  It has been shown that with zero cosmological constant and  zero gravitomagnetic  monopole all these spacetimes  must have the gyromagnetic ratio $ g = 2 $. It should be noted that the presence of an additional dilaton and other fields in four dimensions  makes the gyromagnetic ratio different from $ g = 2 $ (see, for instance \cite{dean}). The behavior of the gyromagnetic ratio in higher-dimensional spacetimes has been the subject of recent studies in \cite{af, aliev1}.

It is worth to emphasize that the most crucial point in all cases described above is that the gyromagnetic ratio has been studied in spacetimes with asymptotic flat structure. Therefore, it is a natural question to ask how does the gyromagnetic ratio change in rotating spacetimes with {\it asymptotic non-flat structure}? In recent works \cite{aliev3, aliev4} this question was examined for rotating charged black holes in four and higher dimensions with asymptotic anti-de Sitter (AdS) structure. It was shown that in four dimensions the Kerr-Newman-AdS black holes must have the same gyromagnetic ratio  as the usual Kerr-Newman black holes. This extends the result of  \cite{gair} for Carter  separable spacetimes to include a non-zero cosmological constant. Furthermore, an interesting behavior of the gyromagnetic ratio was found for charged Kerr-AdS black holes with a single angular momentum  in all higher dimensions. It turned out that in the limit of maximal rotation determined by the curvature radius of the AdS spacetime,  $ g \rightarrow 2 $ regardless of the spacetime dimension. This result was further confirmed in \cite{kunz} within numerical calculations.

In this paper we continue the study of the gyromagnetic ratio for rotating spacetimes with asymptotic non-flat structure. We begin with the most instructive case of a Kerr-Newman black hole pierced by a straight cosmic string. A model of this kind  was first considered in \cite{ag}. The presence of the cosmic string on the symmetry axis of the black hole will destroy the asymptotic structure of the metric which becomes  asymptotically conical rather than asymptotically flat one. Another example of destroying  the asymptotic structure of the Kerr metric is realized in the familiar Kerr-Taub-NUT solution of the Einstein field equations \cite{demianski}. This solution describes a localized axisymmetric source which,  in addition to the ordinary mass
(gravitoelectric charge),  also carries  a gravitomagnetic charge (NUT charge). It is the NUT charge that destroys the asymptotic flat structure of the metric and in this sense it may be thought of as a "measure" of the asymptotic non-flatness. The different aspects of the Kerr-Taub-NUT family of solutions were investigated  by many authors (see \cite{dadhich, nouri, bini1,  bini2} and references therein). We also consider a general class of four-dimensional metrics which involve both cosmological constant and gravitomagnetic monopole charge and for which the Hamilton-Jacobi and wave equations admit separation of variables. As a higher-dimensional example of rotating spacetimes with asymptotic non-flat structure, we examine recently-discovered general black hole solution of minimal five-dimensional supergravity with a cosmological constant \cite{cclp}.

The organization of the paper is as follows. In Sec.II we discuss properties of the Kerr-Newman spacetime threaded by a cosmic string. We evaluate the physical parameters of the spacetime: The charge, mass, and angular momentum. Here we also determine the magnetic dipole moment and show that the gyromagnetic ratio is not affected by the conical parameter of the spacetime.  In Sec.III we  consider Kerr-Newman-Taub-NUT spacetime and describe the effect of the NUT charge on the rotational and electromagnetic properties of the spacetime metric. We introduce two equivalent definitions for the gyromagnetic ratio of this spacetime  and  find that it must have the value $ g=2 $. We also consider a general class of Carter separable spacetimes and  show that they all possess the  same gyromagnetic ratio $ g=2 $ independent of their asymptotic nature. In Sec.IV we study the gyromagnetic ratio of general rotating black holes in minimal five-dimensional gauged  supergravity. We obtain expressions for two independent gyromagnetic ratios associated with two orthogonal 2-planes of rotation. We also discuss some interesting special cases of these expressions.

\section{Kerr-Newman black holes pierced by a cosmic string}

The spacetime around a rotating and charged black hole containing a straight cosmic string on its axis of symmetry is described by the Kerr-Newman solution with asymptotically conical behavior . It is given by the metric
\begin{eqnarray}
ds^2 & = & -{\Delta\over \Sigma}\,(\,dt - a\,b_0 \sin^2\theta\,d\phi\,)^2 + \Sigma \left(\frac{dr^2}{\Delta} +
d\theta^{\,2}\right) \nonumber\\[2mm] &&
+\,{\sin^2\theta\over \Sigma}\,
\left[a \, dt - (r^2+a^2)\, b_0 d\phi \right]^2 \,\,,
\label{knconical}
\end{eqnarray}
where
\begin{eqnarray}
\Delta &= & r^2 - 2M r + a^2 +Q^2 \,,~~\Sigma = r^2 +
 a^2 \cos^2\theta\,\,.
\label{metfunct1}
\end{eqnarray}
The determinant of the metric  is given by
\begin{eqnarray}
\sqrt{-g}  &= & b_{0} \Sigma \sin\theta\,\,.
\end{eqnarray}
In the absence of the cosmic string, $ M $ is  the mass, $ a=J/M $ is the rotation parameter, and  $ Q $ is  the electric charge of the black hole. The parameter $ b_0 = 1-4 \mu $   accounts for physical effects of the cosmic string and makes the asymptotic behavior of the metric conical. Here  $\mu $ is a linear mass density, $ 0< b_0 \leq 1 $ and we use the units in which $ c=\hbar=G=1 $. It can be easily checked that the asymptotic limit of the metric (\ref{knconical}), after an appropriate coordinate transformation, is equivalent to the canonical form of the metric outside of a straight  cosmic string \cite {hiscock}
\begin{eqnarray}
ds^2 &=& - dt^2 + d\rho^2 + dz^2 + b_0 ^2 \rho^2 d\varphi^2\,\,,
\label{stringm}
\end{eqnarray}
where $ (\rho, z, \varphi) $ are the usual  cylindrical coordinates with  $  0 \leq \varphi< 2\pi $.

The potential one-form that describes the corresponding electromagnetic field in the spacetime (\ref{knconical}) is given by
\begin{equation}
A = -{Qr\over \Sigma}\,\left(dt - a b_0 \sin^2\theta\,d\phi \right)\,
\label{kspot}
\end{equation}
and for the nonvanishing covariant components of the electromagnetic field tensor,  we have
\begin{eqnarray}
F_{\,01}& = & {Q\over \Sigma^2}\,(\Sigma- 2 r^2)\,, ~F_{\,13}
= {Q a b_0\over \Sigma^2}\,(\Sigma- 2 r^2)\sin^2\theta\,,\nonumber
\\[2mm]
F_{\,02}& = & {Q r a^2 \sin2\theta\over \Sigma^2}\,,~
F_{\,23} = {Q r a b_0 \over \Sigma^2}\,(r^2 + a^2)\sin2\theta\,.
\label{ftensor}
\end{eqnarray}
In the following we shall also need contravariant components of this tensor, for which  we obtain
\begin{eqnarray}
F^{\,01}& = &-\frac{Q(r^2 + a^2)}{\Sigma^3}(\Sigma- 2
r^2)\,\,,~~F^{\,13} = {Q a \over
{\Sigma^3  b_0}}\,(\Sigma- 2 r^2)\,\,,\nonumber \\[2mm]
F^{\,02}&=& -{Q r a^2 \over
\Sigma^3}\,{\sin2\theta}\,\,,~~~~~~~ F^{\,23} = {2
Q a r\over {\Sigma^3  b_0}}\,\cot\theta\,.
\label{contra}
\end{eqnarray}
It is clear that the physical parameters of the  metric (\ref{knconical}) containing a cosmic string will differ from those of the usual Kerr-Newman metric. Indeed, evaluating the Gaussian flux for the electric charge
\begin{equation}
Q^{\,\prime}= {\frac{1}{4 \pi}} \oint \,^{\star}F\,\,
\label{gauss}
\end{equation}
we find that
\begin{equation}
Q^{\,\prime}=  Q b_0 \,.\label{physcharge1}
\end{equation}
We note that here and in what follows $ F=dA $, the star
stands for the Hodge dual and the prime refers to physical quantities.
The existence of two commuting Killing vector $~ \xi_{(t)}= \partial_t ~$ and $~\xi_{(\phi)}= \partial_\phi \,$ in the spacetime (\ref{knconical})  enables us to use the standard Komar formulas for the mass and angular momentum
\begin{eqnarray}
M^{\prime} & = & -\frac{1}{8 \pi\,}\oint \,^{\star}d\hat
\xi_{(t)}\,,~~~
J^{\prime} =  \frac{1}{16 \pi\,}\oint \,^{\star}d\hat
\xi_{(\phi)}\,,
\label{komars}
\end{eqnarray}
where the Killing one-form $\,\hat \xi=\xi_{\mu}\, d x^{\mu}$ and the integration is performed over a  two-dimensional hypersurface at spatial infinity. The calculations yield
\begin{eqnarray}
M^{\prime} & = & M b_0 \,,~~~~~ J^{\prime} =   J b_0^2\,\,.
\label{massangm0}
\end{eqnarray}
The cosmic string also changes the angular velocity and the surface area of the horizon. We find that
\begin{eqnarray}
\Omega^{\prime}_{H} &=& \frac{a}{b_0\,( r_{+}^2 + a^2)}\,\,,~~~
A^{\prime} = 4 \pi b_0 (r_{+}^2 + a^2)\,\,,
\label{surfa}
\end{eqnarray}
where $ r_{+} $ is the radius of the horizon determined by the equation $\Delta=0 $. On the other hand, it is easy to show that the electrostatic potential of the horizon and its Hawking temperature remain unchanged. That is, we have the usual expressions
\begin{eqnarray}
\Phi_{H} &=& \frac{ Q r_{+}}{ r_{+}^2 + a^2}\,\,,~~~
T  = \frac{r_{+}-M}{2\pi (r_{+}^2 + a^2)}\,\,.
\label{temperature}
\end{eqnarray}
It is straightforward to  verify  that the above quantities satisfy the first law of thermodynamics
\begin{equation}
dE = TdS  +  \Omega^{\prime}_{H} dJ^{\prime} + \Phi_{H}d Q^{\prime} \,,
\label{firstlaw}
\end{equation}
where the total energy $ E=M^{\prime} $ and the entropy $ S= A^{\prime}/4 $.

Next, we  calculate the magnetic dipole moment. We first define an orthonormal frame given  by the basis one-forms of the metric (\ref{knconical})
\begin{eqnarray}
\label{basis1}
e^{0} &=&\left(\frac{\Delta}{\Sigma}\right)^{1/2} \left(dt-a b_0 \sin^2\theta\,d\phi
\right) \,, \nonumber \\[2mm]
e^{3} &=&\frac{\sin\theta}{\Sigma^{1/2}} \left[a \,dt- b_0 (r^2+a^2)\,d\phi
\right]\,\,, \\[2mm]
 e^{1} &=&\left(\frac{\Sigma}{\Delta
}\right)^{1/2}dr\,\,,~~~e^{2}=
\Sigma^{1/2} d\theta\,. \nonumber
\end{eqnarray}
The electromagnetic field two-form written in this frame takes the form
\begin{equation}
F= \frac{Q}{\Sigma^2} \left[ \left(\Sigma- 2 r^2 \right) \,e^{0}\wedge e^{1}
 + 2 r a \cos\theta \,e^{3}\wedge e^{2} \right]\,\,.
 \label{emf1}
\end{equation}
Thus, an observer in the frame (\ref{basis1}) sees only the radial components of the electric and magnetic fields. For the leading behavior of these fields at spatial infinity, we have
\begin{eqnarray}
\label{eradial}
E_{\,\hat r} &=& \frac{Q^{\,\prime}}{b_0 r^2}
+\mathcal{O}\left(\frac{1}{r^4}\right) \,\,,\\[3mm]
B_{\,\hat r} &=& \frac{2 Q^{\,\prime} a
}{b_{0} r^3}\,\cos\theta +\mathcal{O}\left(\frac{1}{r^5}\right)
\,\,.\label{mradial}
\end{eqnarray}
It follows that $ Q^{\,\prime} $ is  the electric charge as given in (\ref{physcharge1}) and the radial magnetic field is determined by the magnetic dipole moment
\begin{eqnarray}
\mu^{\,\prime}&=&  Q a b_{0}^2= \mu b_{0}^2\,.
\label{magd}
\end{eqnarray}
Defining now the gyromagnetic ratio by the relation
\begin{equation}
\mu^{\,\prime}= g\, \frac{ Q^{\,\prime}J^{\,\prime}}{2
M^{\,\prime}}\,\, \label{g}
\end{equation}
we have $ g=2 $.  We conclude that  the gyromagnetic ratio of the Kerr-Newman black hole  with asymptotic conical structure remains equal to $ 2 $, thus being not affected by the cosmic string on its axis of symmetry.

It may be useful to note that in the small-charge limit, $ Q \ll M \,, $ we can also prove that $ g=2 $  independent of the conical parameter $ b_0 $. For this purpose, we employ an approach which is
similar to that of \cite{wald}. (See also Refs.\cite{af, aliev1, aliev3, aliev4}). We begin with the twist one-form
\begin{equation}
{\hat \omega}  =
\frac{1}{2}\,^{\star}\left({\hat\xi}_{(t)} \wedge
d\,{\hat\xi}_{(t)}\right)\,\,. \label{ctwist1}
\end{equation}
Writing out this quantity explicitly in the metric (\ref{knconical}) with $\, Q \ll M\, $, we obtain
\begin{equation}
\hat \omega  = - \frac{a M}{\Sigma^2}\left[2 r \cos\theta \,dr  - (\Sigma - 2 r^2) \sin\theta \,d\theta \right]\,\,.
\label{ctwist2}
\end{equation}
It is easy to check that this equation implies the existence of a scalar twist potential given by
\begin{equation}
\Omega = {M a \cos\theta\over \Sigma}\,.
\label{cstwist3}
\end{equation}

Now we define the magnetic field one-form
\begin{eqnarray}
{\hat B} &=& i_{\hat\xi_{(t)}}\, ^{\star}F =\,
^{\star}\left({\hat\xi_{(t)}}\wedge F\right)\,\,, \label{mform1}
\end{eqnarray}
which written out explicitly, by making use of the expressions in (\ref{contra}), takes the form
\begin{equation}
\hat B =  \frac{a Q}{\Sigma^2}\left[2 r \cos\theta \,dr  - (\Sigma - 2 r^2) \sin\theta \,d\theta \right]\,\,.
\label{mform2}
\end{equation}
This expression can also be put in the form
\begin{equation}
{\hat B}= -d\,\Psi \,\,, \label{magpot}
\end{equation}
where the  magnetostatic potential is given by
\begin{equation}
\Psi = {Q a \cos\theta\over \Sigma\,}\,. \label{cmagpot3}
\end{equation}
It is important to note that the expressions in (\ref{ctwist2}), (\ref{cstwist3}) and (\ref{mform2}), (\ref{cmagpot3})  do not contain the conical parameter $\, b_0 \, $ of the spacetime  at all.

Comparison of equations (\ref{cstwist3}) and  (\ref{cmagpot3}) leads to the relation
\begin{equation}
\Psi = {Q\over M}\,\Omega\,\,
 \label{grelation}
\end{equation}
that can be used, on equal footing with equation (\ref{g}), as a new definition for the gyromagnetic ratio. This proves that $ g=2 $.

\section{Kerr-Newman-Taub-NUT Spacetime}

The Kerr-Newman-Taub-NUT spacetime is  one of the most instructive examples of spacetimes with asymptotic non-flat structure known
in general relativity \cite{demianski}. It is  a stationary axially symmetric solution of the Einstein-Maxwell equations which describes a rotating electrically  charged source  endowed with NUT charge as well. The corresponding metric of the spacetime  can be written in the form
\begin{eqnarray}
ds^2 & = & -{{\Delta}\over {\Sigma}} \left(\,dt - \chi
\,d\phi\,\right)^2 + \Sigma \left(\frac{dr^2}{\Delta}\,+\,
d\theta^{\,2}\right) \nonumber\\[2mm] &&
+ \,\frac{\sin^2\theta}{\Sigma} \left[a\, dt -
\left(r^2+a^2+\ell^2\right) d\phi\, \right]^2 , \label{4kntnut}
\end{eqnarray}
where
\begin{eqnarray}
\label{metfunct}
\Delta &= & r^2 - 2M r + a^2 - \ell^2 +Q^2 \,\,, \nonumber\\[2mm]
\Sigma &= & r^2 + (\ell + a \cos\theta)^2 \,\,,\\[2mm]
\chi & =&  a \sin^2\theta - 2\,\ell \cos\theta \,\,. \nonumber
\end{eqnarray}
The parameter $ M $  is the gravitational mass of the source, $ a $  is its rotation parameter, $ Q $  and  $ \ell $  are the electric  and NUT charges respectively. The determinant of the metric is given by
\begin{equation}
\sqrt{-g} = \Sigma \sin\theta\,
\end{equation}
and for the inverse metric components, we have
\begin{eqnarray}
g^{\,00} & = & -{1\over \Delta\Sigma}\,[\,(r^2 + a^2 + \ell^2)^2 - \Delta\,a^2 \sin^2\theta \nonumber \\ &&
+\,4 \ell \,\Delta {\cot\theta\over \sin\theta}\, (\chi +
\ell \cos\theta)\,]\,,~~~~ g^{\,11} = {\Delta\over \Sigma}\,\,,
\nonumber \\[2mm]
g^{\,22} & = & {1\over
\Sigma}\,,~~~~ g^{\,33} = {1\over \Delta\Sigma\sin^2\theta}\,
(\Delta - a^2\sin^2\theta)\,,\\[2mm]
g^{\,03} &=& -{1\over \Delta\Sigma}\,\left[\,(2Mr+ 2\,\ell^2- Q^2 )\,a
+ 2\ell \,\Delta\,{\cot\theta\over \sin\theta}\,\right]\,.\nonumber
\end{eqnarray}

The electromagnetic field of the source is described by the potential one-form
\begin{equation}
A= -\frac{Q \,r}{\Sigma}\left(dt-  \chi\,d\phi \right)\,\,,
\label{potform}
\end{equation}
leading to the following non-vanishing components of the electromagnetic field tensor
\begin{eqnarray}
\label{emt}
F_{\,01} &= &{Q\over \Sigma^2}\,(\Sigma- 2 r^2)\,,~~~~
F_{\,13} =  {Q\,\chi\over \Sigma^2}\,(\Sigma-2 r^2)\,,\nonumber
\\[2mm]
F_{\,02}& = &{2\,Q r a \over \Sigma^2}\,(\ell + a\cos\theta)\,\sin\theta\,, \\[2mm]
F_{\,23} &= & {2\,Q r\over \Sigma^2}\,\sin\theta \,(r^2 + a^2 +
\ell^2)\,(\ell + a\,\cos\theta)\,. \nonumber
\end{eqnarray}
The contravariant components of this tensor  are given by
\begin{eqnarray}
\label{contraemt}
F^{\,01}& = &-{Q \over \Sigma^3}\,(r^2 + a^2 + \ell^2)\,(\Sigma -
2 r^2)\,,\nonumber \\[2mm]
F^{\,02}& = &-{Q\,r\,\over \Sigma^3}\left[a^2\sin2\theta -
4\ell^2\,\cot\theta - {a\ell\over \sin\theta}(1 +
3\cos2\theta)\right],
\nonumber \\[2mm]
F^{\,13} &= & {a Q\over \Sigma^3}\,(\Sigma - 2r^2)\,,~~~
F^{\,23}  =  {2 Q r\over \Sigma^3}\,{(\ell + a\cos\theta)\over
\sin\theta}\,.
\end{eqnarray}
With vanishing NUT charge, $ \ell=0 $, the above expressions  go over into (\ref{ftensor}) and (\ref{contra}) taken for $ b_0=1 $.

Inspection the Komar integrals (\ref{komars}) for the metric (\ref{4kntnut}) at spatial infinity confirms that $M^{\prime}= M $  and $J^{\prime}= J= a M $.  Indeed, for the dominant behavior of the corresponding  integrands, we have
\begin{eqnarray}
\label{expansion1}
\xi_{\,(t)}^{\,t;\,r} & = & {M\over r^2} + \mathcal{O}\left({1\over
r^4}\right)\,,\\[2mm]
\xi_{\,(\varphi)}^{\,t;\,r} &=&
- \frac{3 M\,\chi}{r^2}  - \frac{2\ell\,\cos\theta}{r}
+ \mathcal{O}\left({\,1\over r^4}\right)\,.
\label{expansion2}
\end{eqnarray}
Inserting these expansions into (\ref{komars}) and performing the integration over a two-sphere at infinity we verify the statement made above. We note that the second term in (\ref{expansion2}) makes
no contribution to the surface integral due to the integration over the angle $ \theta $. Thus, the NUT charge does not affect the usual Komar expressions for the mass and angular momentum.

However, the physical effect of the NUT charge can be read off
from the asymptotic behavior of the metric (\ref{4kntnut}). For this purpose, following our previous works \cite{aliev3, aliev4}, we
employ  a ``background subtraction" approach.  This yields
\begin{eqnarray}
\delta g_{03}& = & - \,\frac{2 M (a \sin^2\theta-2 \ell \cos\theta)}{r} + \mathcal{O}\left(\frac{1}{r^{3}}\right), \label{renasympg03}
\end{eqnarray}
where we have taken the difference between the off-diagonal components of the metric under consideration and its reference background. The latter is obtained from the former by setting $ M=0 $ in it.
This expression shows that in addition to the usual  angular momentum or gravitomagnetic dipole moment $ J_{D} $ (in the terminology of gravitomagnetism), the source can  also be assigned a  ``specific angular momentum" $ j_{M} $ of similar order determined by the NUT charge (gravitomagnetic monopole moment). That is, we have
\begin{eqnarray}
J_{D}  & = &  a M \,\,,~~~~~~~~~ j_{M}= \ell M\,\,.
\label{dipmon}
\end{eqnarray}

In accordance with these two angular momenta, the Kerr-Newman-Taub-NUT  source must also have  two magnetic moments. To see this, it is enough to consider the asymptotic form of the associated radial magnetic field. Again, we choose an orthonormal frame
\begin{eqnarray}
\label{basis}
 e^{0} &=&\left(\frac{\Delta }{\Sigma
}\right)^{1/2}\left(dt- \chi\,d\phi
\right) \,, \nonumber \\[3mm]
e^{3} &=&\frac{\sin\theta}{\Sigma^{1/2}}\left[a\, dt -
\left(r^2+a^2+\ell^2\right) d\phi\, \right]\,\,, \\[3mm]
 e^{1} &=&\left(\frac{\Sigma}{\Delta
}\right)^{1/2}dr\,\,,~~~~~e^{2}= \Sigma^{1/2} d\theta \,,
\nonumber
\end{eqnarray}
in which, for the radial electric and magnetic fields, we obtain
\begin{eqnarray}
\label{eradials}
 E_{\,\hat r} &=&\frac{Q}{r^2}
+\mathcal{O}\left(\frac{1}{r^4}\right) \,\,,\\[3mm]
 B_{\,\hat r} &=&\frac{2 Q \left(\ell+ a \cos\theta \right)}{r^3} +\mathcal{O}\left(\frac{1}{r^5}\right)
\,\,.\label{mradials}
\end{eqnarray}
The first expression is the Coulomb  field of the electric charge $ Q $ . From the second expression it follows that the radial magnetic field is generated by  two independent quantity; the usual magnetic dipole moment $ \mu_{D} $  and  the  "specific magnetic moment" $ \mu_{M} $ which is due to the NUT charge. Thus, we have
\begin{eqnarray}
\mu_{D}  & = &  a Q\,\,,~~~~~~~~~ \mu_{M}= \ell Q \,\,.
\label{mdipmon}
\end{eqnarray}
These considerations allow us to write down the following defining relations for the gyromagnetic ratio
\begin{eqnarray}
\mu_{D}  & = & g_1\, \frac{Q J_D}{2 \,M}\,\,,~~~~ \mu_{M}= g_2\, \frac{Q j_M}{2 \,M}\,\,.
\label{defgyros}
\end{eqnarray}
Comparing the corresponding expressions in (\ref{dipmon}) and (\ref{mdipmon}), we find that $ g_1=g_2=2 $. Hence, the gyromagnetic ratio for a Kerr-Newman-Taub-NUT  source endowed with both electric and NUT charges is $ g=2  $.  We see that this value remains unchanged, when the rotation parameter of the source vanishes.

As in the previous section, one can prove the above value of the gyromagnetic ratio employing the formalism of twist potentials.  Again assuming small electric charge, $ Q\ll M $,  and calculating the twist one-form in (\ref{ctwist1}) for the metric (\ref{4kntnut}), we obtain
\begin{eqnarray}
\hat \omega & = & \frac{a \sin\theta }{\Sigma^2} \left[M (\Sigma -2 r^2)- 2 r \ell (\ell+a \cos\theta)\right] d\theta \nonumber \\[2mm]  & &
- \,\frac{2 M r(\ell+a \cos\theta) +\ell ( \Sigma -2 r^2)}{\Sigma^2}\, dr
\label{nuttwist}
\end{eqnarray}
Straightforward calculations show that the potential one-form is generated by the scalar twist
\begin{equation}
\Omega = {a\,M\cos\theta + \ell\,(\,M-r)\over \Sigma}\,.
\label{stwist}
\end{equation}
This expression involves an undesired term that is not vanishing  for $ M=0$. This is  due to the asymptotic non-flat structure of the spacetime under consideration. To make the scalar twist physically meaningful, we perform the background subtraction which yields
\begin{equation}
\delta \Omega =  \frac{M(\ell+a \cos\theta)}{\Sigma}
\label{restwist}
\end{equation}
It follows that for zero rotation parameter $ a $  of the metric (\ref{4kntnut}), the timelike Killing vector $\partial_{t} $ still fails to be hypersurface orthogonal, that is the NUT charge generates an additional "rotation" of the metric.

Similarly, using the expressions in (\ref{contraemt}) one can
evaluate the magnetic field one-form (\ref{mform1}). We find that
\begin{equation}
\hat B = { Q \over \Sigma^2}\,\left[2\, r(\ell+a\,cos\theta\,) \,dr
- a\,(\,\Sigma - 2r^2\,)\sin\theta \, d\theta \right] .
\end{equation}
This allows us to introduce a magnetostatic potential of the form
\begin{equation}
\Psi = {Q\over \Sigma}\,(\ell+a\,cos\theta\,)\,. \label{nutmagpot}
\end{equation}
This equation along with  that in (\ref{restwist}) leads us to the relation
\begin{equation}
\Psi = {Q\over M}\,\delta \Omega\,
 \label{tmrelation}
\end{equation}
that is equivalent to the definitions in (\ref{defgyros}). In this respect, it remarkably combines them into one, thereby proving the gyromagnetic ratio $ g=2 $ for the Kerr-Newman-Taub-NUT source.

\subsection{Inclusion of the Cosmological Constant}

To complete the above discussion of the gyromagnetic ratio for rotating spacetimes with asymptotic non-flat structure in four dimensions, we consider now a general class of metrics which admit separation of variables for the Hamilton-Jacobi  and Schr\"{o}dinger equations \cite{carter}. As is known, these Carter separable spacetimes  include a cosmological constant as well. Using the Plebanski approach  \cite{plebanski} one can find the most convenient form for these metrics. It is given by
\begin{eqnarray}
ds^2 & = & -{{Y}\over {r^2 +p^2}} \left( d\tau - p^2
\,d\psi\,\right)^2 +{{r^2+p^2}\over {Y}}\,dr^2   \nonumber\\[2mm] &&
+\,{{r^2+p^2}\over {X}}\,dp^2 + {{X}\over {r^2 +p^2}} \left( d\tau + r^2\,d\psi\,\right)^2,
\label{gmetric}
\end{eqnarray}
where
\begin{eqnarray}
X&= & \gamma  - \epsilon\, p^2 - \lambda  p^4 + 2 L p \,\,, \nonumber\\[2mm]
Y &= & \gamma + \epsilon\, r^2 - \lambda\, r^4 - 2 M r +Q^2\,\,.
\label{gmetfunct}
\end{eqnarray}
The parameters $\, M\,, \lambda \,, L\,,Q \, $  are related to the mass, cosmological constant, NUT  and electric charges respectively.  The remaining  $\,\gamma \,$  and $\,\epsilon \,$  are arbitrary real parameters. We have set the possible  magnetic charge equal to zero.

The potential one-form for the electromagnetic field of these metrics is given by
\begin{equation}
A= -\frac{Q \,r}{{r^2 +p^2}}\left(d\tau- p^2 \,d\psi \right)\,\,.
\label{gpotform}
\end{equation}
These metrics satisfy the coupled system of the Einstein-Maxwell equations with the cosmological constant $ \Lambda= 3 \lambda $.

If we choose the parameters
\begin{eqnarray}
\gamma &= & \left(a^2-\ell^2 \right)\left(1+ \lambda \,\ell^2\right)\,,~~
\epsilon= 1 - \lambda \left(a^2+ 2 \ell^2 \right)\,,
\nonumber\\[2mm] &&
L = \ell\,\left(1+ \lambda a^2\right) \,\,,
\label{parameters}
\end{eqnarray}
and make  the  redefinitions of the coordinates  given by
\begin{eqnarray}
\tau &= & t-\frac{a^2+\ell^2}{a\, \Xi} \,\phi\, ,~~~~
\psi= - \frac{\phi}{a \,\Xi} \,,
\nonumber\\[2mm] &&
p= \ell+ a \cos\theta\,\,,
\label{recoord}
\end{eqnarray}
where $ \Xi = 1+ \lambda\, a^2 \,,$  the metric (\ref{gmetric}) takes the form
\begin{eqnarray}
ds^2 & = & -{{\Delta_r}\over {\Sigma}} \left(\,dt - \frac{\chi}{\Xi}
\,d\phi\,\right)^2 + \Sigma \left(\frac{dr^2}{\Delta_r}\,+\,
\frac{d\theta^{\,2}}{\Delta_{\theta}}
\right) \nonumber\\[2mm] &&
+ \,\frac{\Delta_{\theta} \sin^2\theta}{\Sigma} \left[a\, dt -
\frac{r^2+a^2+\ell^2}{\Xi} d\phi\, \right]^2 \,, \label{gmbl}
\end{eqnarray}
where
\begin{eqnarray}
\Delta_r &= & \left(r^2 + a^2 - \ell^2 \right) \left(1 -\lambda\, r^2 \right)  - 2 M r +Q^2 \nonumber\\[1mm] &&
+ \,\lambda \ell^2 \left(a^2-\ell^2 - 3 r^2\right)
\,\,, \\[2mm]
\Delta_{\theta} &= & 1 + \lambda \,\left(2 \ell + a \cos\theta\right)^2 \,\,
\label{gmetfunct1}
\end{eqnarray}
and the metric functions $ \Sigma $ and $ \chi $ are the same as those given in (\ref{metfunct}). (See also \cite{podolsky} for an alternative  choice of the parameters). As for the potential one-form (\ref{gpotform}), it  reduces to
\begin{equation}
A= -\frac{Q \,r}{\Sigma}\left(dt- \frac{\chi}{\Xi} \,d\phi \right)\,\,.
\label{gpotform1}
\end{equation}
The metric (\ref{gmbl}) is a generalization of  the Kerr-Newman-Taub-NUT solution (\ref{4kntnut}) to include the cosmological constant. In what follows, we take the cosmological constant to be negative, without loss of generality.

For the electromagnetic field two-form we obtain
\begin{eqnarray}
\label{2form}
 F&=& \frac{Q \left(\Sigma- 2 r^2 \right)
}{\Sigma{^2}} \left(dt-\frac{\chi}{\Xi}\, d\phi
\right)\wedge dr \,+\nonumber\\[1mm] &&
\frac{2 Q r (\ell+a \cos\theta) \sin \theta}{\Sigma^2} \left(a dt - \frac{r^2+a^2+\ell^2}{\Xi} d\phi \right)
\wedge d\theta . \nonumber\\
\end{eqnarray}
The contravariant components of the electromagnetic field  tensor $ F^{01} $ and $ F^{02} $  are the same as those given in (\ref{contraemt}), while the remaining components $ F^{13} $ and $ F^{23} $ are equal to their counterparts in (\ref{contraemt}) multiplied by $ \Xi $.

The electric charge of the  source (\ref{gmbl}) is given by the Gaussian flux in (\ref{gauss}), which now yields
\begin{equation}
Q^{\,\prime}=  \frac{Q}{\Xi} \,\,.\label{charge}
\end{equation}

The physical mass and angular momentum  can be evaluated following our previous work \cite{aliev4}. Performing the similar calculations for the spacetime (\ref{gmbl}) we obtain that
\begin{eqnarray}
M^{\prime}&=& \frac{M}{\Xi^2}\,\,,~~~~~~ J^{\prime}_{D}= \frac{a
M}{\Xi^2}\,\,\label{massangm}
\end{eqnarray}
that is, the same expressions as for the Kerr-Newman-AdS spacetime \cite{gpp1}. Again, the effect of the NUT charge can be seen from
the asymptotic behavior of the metric.  Performing   an appropriate background subtraction, as in the case of  (\ref{renasympg03}), we find  "the specific angular momentum"  due to the NUT charge
\begin{eqnarray}
j^{\,\prime}_{M}= \frac{\ell M}{\Xi^2}.
\label{gdipmon}
\end{eqnarray}
We note that this quantity  and the angular momentum in (\ref{massangm})  generalize those  given in (\ref {dipmon}) to include  the case of a negative cosmological constant.

Next, evaluating the orthonormal  components of the radial electric  and magnetic fields, we find the expressions similar to (\ref{eradials}) and (\ref{mradials})
\begin{eqnarray}
\label{geradials}
 E_{\,\hat r} &=&\frac{Q^{\prime}\,\Xi}{r^2}
+\mathcal{O}\left(\frac{1}{r^4}\right) \,\,,\\[3mm]
 B_{\,\hat r} &=&\frac{2 Q^{\prime}\,\Xi \left(\ell+ a \cos\theta \right)}{r^3} +\mathcal{O}\left(\frac{1}{r^5}\right)
\,\,.\label{gmradials}
\end{eqnarray}
The first expression gives the electric charge in (\ref{charge}), while the second expression determines the  counterparts of the magnetic momenta in (\ref{mdipmon}) for the Kerr-Newman-Taub-NUT-AdS spacetime. We have
\begin{eqnarray}
\mu^{\prime}_{D}  & = &  \frac{a Q}{\Xi}\,\,,~~~~~~~~~ \mu^{\prime}_{M}= \frac{\ell Q}{\Xi} \,\,.
\label{gmdipmon}
\end{eqnarray}
It is now easy to see that one can define the gyromagnetic ratio in terms of the primed quantities as follows
\begin{eqnarray}
\mu^{\prime}_{D}  & = & g_1\, \frac{Q^{\prime} J^{\prime}_D}{2 \,M^{\prime}}\,\,,~~~~ \mu^{\prime}_{M}= g_2\, \frac{Q^{\prime} j^{\prime}_M}{2 \,M^{\prime}}\,.
\label{prdefgyros}
\end{eqnarray}
It follows immediately that  $ g_1=g_2=2 $. This result shows that all Carter separable spacetimes satisfying the coupled system of the Einstein-Maxwell equations in four dimensions must have the gyromagnetic ratio $ g=2 $ independent of their asymptotic nature.

\section{Rotating charged black holes in minimal Five-dimensional  gauged supergravity}

In this section we wish to consider an example of a higher-dimensional rotating charged spacetime with asymptotic non-flat structure. For this purpose, we take the general solution for rotating charged black holes in minimal five-dimensional  gauged supergravity recently found  in \cite{cclp}  by Chong, Cveti\u{c}, Lu and Pope (CCLP). The CCLP black holes  are determined by a set of  four  conserved charges: The mass, electric charge  and two angular momenta. There are several motivations to calculate the gyromagnetic ratio of these black holes.  First of all, they  possess  asymptotic non-flat, AdS/dS  structure and  the electric charge may have arbitrary values up to the BPS limit. This allows  to study  the combined influence of  both asymptotic structure and the electric charge on the gyromagnetic ratio of these black holes.

In a recent work \cite{aliev4} we calculated the gyromagnetic ratio for general five-dimensional Kerr-AdS  black holes  carrying  small (test) electric charge. Since the exact metric for the charged Kerr-AdS  black holes in  the Einstein-Maxwell theory is still absent (see \cite{aliev1},\cite{aliev2}), the question of how  our result will change for generic values of the electric charge remains unclear. Though the CCLP  spacetime  is not the same as that needed for the Einstein-Maxwell case, it may  nevertheless  serve as  a good example to evaluate the gyromagnetic ratio for generic values of the electric charge.

Finally, the rotating black hole solutions in AdS background are of interest in the context of probing AdS/CFT correspondence. In \cite{hhtr} it has been shown that there is a deep similarity between the thermodynamic properties of the five-dimensional Kerr-AdS  black holes and their CFT (conformal field theory) dual on the boundary Einstein space rotating at the speed of light. In this respect, one may hope that the calculation of the gyromagnetic ratio for the rotating AdS  black holes in minimal five-dimensional  gauged supergravity  may have a relevance for a further probing AdS/CFT correspondence in the spirit of \cite{hhtr}.

The CCLP metric satisfies the equation of motion for minimal   five-dimensional gauged supergravity derived from the Lagrangian
\begin{eqnarray}
{\cal L} &=& (R+12/l^2)\,{\star} \,1 -\frac{1}{2} \,\, ^{\star}F\,\wedge F \nonumber \\ &&
+\frac{1}{3\sqrt{3}}\,\, F\,\wedge F\,\wedge A\,\,,
\label{5sugralag}
\end{eqnarray}
where the radius of the AdS background is given by the negative cosmological constant, $ l^2= - 6/\Lambda $. In its original form this metric was given the  Boyer-Lindquist type coordinates $ x^{\mu}=\{t, r, \theta, \phi, \psi\}$ which are asymptotically static \cite{cclp}. For our purposes here, it is useful to employ a similar type of coordinates which are rotating at infinity. In these new coordinates, the metric takes the form
\begin{eqnarray}
\label{gsugrabh}
ds^2 & = & - \left( dt - \frac{a
\sin^2\theta}{\Xi_a}\,d\phi - \frac{b
\cos^2\theta}{\Xi_b}\,d\psi \right)\nonumber
\\ &&
\left[f \left( dt - \frac{a
\sin^2\theta}{\Xi_a}\,d\phi\, - \frac{b
\cos^2\theta}{\Xi_b}\,d\psi \right)\nonumber
\right. \\[2mm]  & & \left. \nonumber
+ \frac{2 Q}{\Sigma}\left(\frac{b
\sin^2\theta}{\Xi_a}\,d\phi + \frac{a
\cos^2\theta}{\Xi_b}\,d\psi \right) \right]
\nonumber \\[2mm] &&
+ \,\Sigma
\left(\frac{dr^2}{\Delta_r} + \frac{d\theta^{\,2}}{
~\Delta_{\theta}}\right) \\[2mm] &&
+ \,\frac{\Delta_{\theta}\sin^2\theta}{\Sigma} \left(a\, dt -
\frac{r^2+a^2}{\Xi_a} \,d\phi \right)^2 \nonumber \\[2mm] &&
+\,\frac{\Delta_{\theta}\cos^2\theta}{\Sigma} \left(b\, dt -
\frac{r^2+b^2}{\Xi_b} \,d\psi \right)^2 \nonumber
\\[2mm] &&
+\,\frac{1+r^2\,l^{-2}}{r^2 \Sigma } \left( a  b \,dt - \frac{b
(r^2+a^2) \sin^2\theta}{\Xi_a}\,d\phi \nonumber
\right. \\[2mm]  & & \left. \nonumber
- \, \frac{a (r^2+b^2)
\cos^2\theta}{\Xi_b}\,d\psi \right)^2,
\end{eqnarray}
where
\begin{eqnarray}
\Delta_r &= &\frac{\left(r^2 + a^2\right)\left(r^2 +
b^2\right)\left(1+r^2l^{-2} \right)+ 2 a b\, Q + Q^2}{r^2} - 2 M  ,\nonumber \\[2mm]
f& =&  {{\Delta_r}\over {\Sigma}} + \frac{Q^2}{\Sigma^2}- \frac{Q^2}{r^2 \Sigma}- \frac{2 a b\, Q}{r^2 \Sigma }\,\,\,, \\[2mm]
\Delta_\theta & = & 1 -\frac{a^2}{l^2} \,\cos^2\theta
-\frac{b^2}{l^2} \,\sin^2\theta \,,~~~~~ \Xi_a=1 -
\frac{a^2}{l^2}\,,\nonumber \\[2mm]
\Sigma & = & r^2+ a^2 \cos^2\theta + b^2 \sin^2\theta \,,~~~~~~
\Xi_b=1 - \frac{b^2}{l^2}\,\,\,. \nonumber
\label{gsugrametfunc}
\end{eqnarray}
We see that the metric is characterized by the mass $ M $, charge $ Q $ and  two independent rotation parameters $ a $ and $ b  $. Furthermore, in the present form it closely resembles the  general Kerr-AdS solution found in \cite{hhtr} covering the latter for $ Q=0 $.

The potential one-form is given by
\begin{equation}
A= -\frac{\sqrt{3}\,Q}{\Sigma}\,\left(dt- \frac{a
\sin^2\theta}{\Xi_a}\,d\phi -\frac{b \cos^2\theta}{\Xi_b}\,d\psi
\right)\,. \label{sugrapotform1}
\end{equation}

In some cases, one also needs to know the  components of the metric (\ref{gsugrabh}). We obtain that
\begin{eqnarray}
g_{00}&=& -1+ \frac{2 M}{\Sigma} -
\frac{r^2 +a^2 \sin^2\theta + b^2 \cos^2 \theta}{l^2}-\frac{Q^2}{\Sigma^2}\,\,, \nonumber \\[2mm]
g_{33}&=& \frac{\sin^2
\theta}{\Xi_{a}^2}\left[\left(r^2+a^2\right) \Xi_a + \frac{2 M a^2
\sin^2 \theta}{\Sigma} \nonumber
\right. \\[3mm]  & & \left. \nonumber
+ \,\frac{a Q \sin^2\theta}{\Sigma^2}\left(2 b  \Sigma - a Q \right)\right], \nonumber \\[2mm]
g_{44}&=& \frac{\cos^2
\theta}{\Xi_{b}^2}\left[\left(r^2+b^2\right) \Xi_b + \frac{2 M b^2
\cos^2 \theta}{\Sigma}
\right. \\[3mm]  & & \left. \nonumber
+ \,\frac{b Q \cos^2\theta}{\Sigma^2}\left(2 a  \Sigma - b Q \right)\right], \nonumber \\[2mm]
g_{03}& = & - \frac{\sin^2
\theta}{\Xi_a}\,\left[a\left(\frac{2 M}{\Sigma}-\frac{r^2+a^2}{l^2}\right)
+\frac{Q}{\Sigma^2}\left( b \Sigma - a Q \right)\right],\nonumber \\[2mm]
g_{04}& = & - \frac{\cos^2
\theta}{\Xi_a}\,\left[b\left(\frac{2 M}{\Sigma}-\frac{r^2+b^2}{l^2}\right)
+\frac{Q}{\Sigma^2}\left( a \Sigma - b Q \right)\right],\nonumber \\[2mm]
g_{34}& = & \frac{\sin^2 \theta\,\cos^2 \theta}{\Sigma^2 \, \Xi_a\, \Xi_b}\,\left[ a b \left(2 M \Sigma-Q^2\right)+ Q \Sigma \left(a^2+b^2\right)\right] \nonumber \\[2mm]
g_{11}& = & \frac{\Sigma}{\Delta_r}\,,~~~~~g_{22}=\frac{\Sigma}{\Delta_{\theta}}
\,.\nonumber \label{sugrametcomp}
\end{eqnarray}
It is interesting to note that the metric determinant does not involve the electric charge. It is given by
\begin{equation}
\sqrt{-g}= \frac{r \Sigma \sin\theta\,\cos\theta}{\Xi_a\,
\Xi_b}\,\,. \label{5ddeterminant}
\end{equation}

The electromagnetic field two-form is given by
\begin{eqnarray}
\label{sugra2form}
 F&=&- \frac{2 \sqrt{3}\,Q  r}{\Sigma^2}\left(dt- \frac{a
\sin^2\theta}{\Xi_a}\,d\phi -\frac{b \cos^2\theta}{\Xi_b}\,d\psi
\right)\wedge dr  \nonumber
\\[2mm] &&
  + \,\frac{\sqrt{3}\,Q a \sin 2\theta}{\Sigma^2} \left(a dt - \frac{r^2+a^2}{\Xi_a}\,d\phi \right)
 \wedge d\theta \nonumber\\[2mm]
& & - \,\frac{ \sqrt{3} \,Q b \sin 2\theta}{\Sigma^2} \left(b dt -
\frac{r^2+b^2}{\Xi_b}\,d\psi \right)
 \wedge d\theta\,\,
\end{eqnarray}
The contravariant components of the electromagnetic field tensor have the form
\begin{eqnarray}
F^{01}&=&\frac{2 \sqrt{3}\,Q\left[a b Q +(r^2+a^2)(r^2+b^2)\right]}{r\,\Sigma\,^3}\,,
\nonumber\\[2mm]
F^{02}&=& \,\frac{ \sqrt{3}\,Q\,(b^2-a^2)\,\sin2\theta}{\Sigma\,^3}\,,
\nonumber \\[2mm]
F^{13}& =& -\, \frac{2 \sqrt{3}\,Q\,\Xi_a \left[Q b + a (r^2+b^2)\right]}{r\,\Sigma\,^3}\,,  \\[2mm]
F^{14}& =& -\, \frac{2 \sqrt{3}\,Q\,\Xi_b \left[Q a + b (r^2+a^2)\right]}{r\,\Sigma\,^3}\,, \nonumber  \\[2mm]
F^{23}& =&\frac{2 \sqrt{3}\,Q\,a\,\Xi_a \cot\theta}{\Sigma\,^3}\,,
\nonumber  \\[2mm]
F^{24} &=&  - \frac{2 \sqrt{3}\,Q\,b\,\Xi_b
\tan\theta}{\Sigma\,^3}\,. \nonumber
\label{Sugraemtcontra}
\end{eqnarray}
Evaluating now the Gauss integral
\begin{equation}
Q^{\,\prime}= {\frac{1}{16 \pi}} \oint \,\left(^{\star}F -F\wedge A/\sqrt{3}\right)\,,
\label{ggauss}
\end{equation}
we find the physical electric charge of the black hole
\begin{equation}
Q^{\,\prime}=\frac{\pi \sqrt{3}\, Q}{4 \, \Xi_a\Xi_b} \,\,.\label{sugracharge}
\end{equation}

The bi-azimuthal symmetry of the spacetime (\ref{gsugrabh}) results in the existence of two commuting Killing vectors  $\xi_{(\phi)}= \partial/\partial \phi\,$ and $\xi_{(\psi)}= \partial/\partial \psi\,$. Thus, we can use the Komar integrals for the angular momenta
\begin{eqnarray}
J^{\prime}_{a}& = & \frac{1}{16 \pi\,}\oint \,^{\star}d\hat
\xi_{(\phi)}\,,~~~J^{\prime}_{b} =  \frac{1}{16 \pi\,}\oint \,^{\star}d\hat \xi_{(\psi)}\,\,.\label{komars1}
\end{eqnarray}
Having performed  straightforward calculations, we obtain
 \begin{eqnarray}
J^{\prime}_{a} &= &\frac{\pi}{4}\,\cdot \frac{2 a M+ b Q (2-\Xi_a)}{\Xi_{a}^2
\Xi_b}\,,\\[2mm]
J^{\prime}_{b} &= &\frac{\pi}{4}\,\cdot \frac{2 b M + a Q (2-\Xi_b)}{\Xi_{a}
\Xi_{b}^2}\,.\nonumber
\label{sugrajj}
 \end{eqnarray}
With these angular momenta, as was shown in \cite{cclp}, the total mass of the black hole satisfying the first law of thermodynamics is given by
\begin{equation}
M^{\prime}=  \frac{\pi M \left(2 \Xi_a +2 \Xi_b-
\Xi_a\ \Xi_b\right) + 2 \pi Q a b \, l^{-2}\left(\Xi_a+\Xi_b\right)}{4\,\Xi_{a}^2 \Xi_{b}^2}
\,\,.\label{sugraphysmass}
\end{equation}

We turn now to the potential one-form in (\ref{sugrapotform1}). It reveals that the black hole must have two magnetic dipole moments associated with two orthogonal 2-planes of rotation. We shall determine these magnetic moments from the asymptotic behavior of the electromagnetic field two-form (\ref{sugra2form}) in an  orthonormal frame. It is remarkable that in the asymptotic region $ r\rightarrow \infty \,$, one can still  use the orthonormal frame constructed in \cite{aliev4}.  Straightforward calculations show that in this frame the leading components of the magnetic field  are given  by the expressions
\begin{eqnarray}
\label{5d asymptotic2}
F_ {\,\hat{2}\hat{3}}&= &\frac{2 \sqrt{3}\, Q a \cos\theta}{r^4}\,\zeta
+\mathcal{O}\left(\frac{1}{r^{6}}\right) \,\,,\\[3mm]
F_ {\,\hat{2}\hat{4}}&= &- \frac{2 \sqrt{3}\, Q b \sin\theta}{r^4}\,\eta
+\mathcal{O}\left(\frac{1}{r^{6}}\right) \,\,,
\label{5d asymptotic3}
\end{eqnarray}
where
\begin{equation}
\label{zeta}
\zeta = \frac{(\Delta_\theta)^{1/2} \left(\Delta_\theta +a^2 l^{-2}\right)^{1/2}+ \left(\Delta_\theta +b^2 l^{-2}\right)^{1/2}}
{(\Delta_\theta)^{1/2}+ \left(\Delta_\theta +a^2 l^{-2}\right)^{1/2}\left(\Delta_\theta +b^2 l^{-2}\right)^{1/2}}\,\,,
\end{equation}
\begin{equation}
\eta = \frac{(\Delta_\theta)^{1/2} \left(\Delta_\theta +b^2 l^{-2}\right)^{1/2}+ \left(\Delta_\theta +a^2 l^{-2}\right)^{1/2}}
{(\Delta_\theta)^{1/2}+ \left(\Delta_\theta +a^2 l^{-2}\right)^{1/2}\left(\Delta_\theta +b^2 l^{-2}\right)^{1/2}}\,\,.
\label{eta}
\end{equation}
We note that for vanishing cosmological constant, $ l\rightarrow \infty \,$, the quantities $ \zeta  $ and  $ \eta  $  tend to unity. They also go to unity for equal rotation parameters. Furthermore, for $ b=0 \,,$ $ \zeta=1 $ or for $ a=0 \,,$ $ \eta=1 $.

From the above expressions it follows that  we can assign two magnetic dipole moments to the black hole which are  given by
\begin{eqnarray}
\mu^{\prime}_{a} & = & a Q^{\prime}= \frac{\mu_a}{\Xi_a \Xi_b} \,,~~~~~~
\mu^{\prime}_{b}  = b Q^{\prime}= \frac{\mu_b}{\Xi_a \Xi_b} \,,
\label{sugradipoles}
\end{eqnarray}
where we have introduced the magnetic moment parameters
\begin{eqnarray}
\mu_{a} & = &  \frac{\pi \sqrt{3}\, Q a}{4} \,,~~~~~~\mu_{b}  =  \frac{\pi \sqrt{3}\, Q b }{4}\,\,.
\end{eqnarray}
The two magnetic dipole moments allow, as discussed earlier in \cite{aliev4}, to define two gyromagnetic ratios  by
\begin{eqnarray}
g_{a} & = & \frac{2 M^{\prime} \mu^{\prime}_{a}}{Q^{\prime} J^{\prime}_{a}} \,,~~~~~~g_{b}  = \frac{2 M^{\prime} \mu^{\prime}_{b}}{Q^{\prime} J^{\prime}_{b}}\,\,.
\label{dipole}
\end{eqnarray}
Substituting into these equations the explicit expressions for the primed quantities, we find
\begin{eqnarray}
g_{a} & = & \frac{2 a}{\Xi_b}\,\cdot \frac{M \left(2 \Xi_a +2 \Xi_b-
\Xi_a \Xi_b\right) +2  Q a b\,l^{-2} \left(\Xi_a+\Xi_b\right)}{2 a M + Q b (2-\Xi_a)}\nonumber\\
\label{gyroa}
\end{eqnarray}
\begin{eqnarray}
g_{b} & = & \frac{2 b}{\Xi_a}\,\cdot \frac{M \left(2 \Xi_a +2 \Xi_b-
\Xi_a \Xi_b\right) +2  Q a b\,l^{-2} \left(\Xi_a+\Xi_b\right)}{2 b M + Q a(2-\Xi_b)}\nonumber\\
\label{gyrob}
\end{eqnarray}
We note that in the small-charge limit, $ Q\ll M $, these expressions agree with those obtained in \cite{aliev4}.

Let us now  consider some special cases of interest;\\
\noindent {i)}~ $ b = 0 ~ $  or $~ a=0 $,
a black hole with a single angular momentum. In this case the gyromagnetic ratio, as follows from the above expressions, is given   either by
\begin{equation}
g_a =  2 + \Xi_a \,,~~~or~~~ g_b =  2 + \Xi_b \,,
\label{singlegyro}
\end{equation}
that is, it does not depend on the electric charge at all. This is the previously known result of \cite{aliev3}.\\
\noindent {ii)}~ $ a = b $\,,~~$ \Xi_a=\Xi_b = \Xi,$ two equal angular momenta. The two gyromagnetic ratios (\ref{gyroa}) and  (\ref{gyrob}) merge into one to give
\begin{equation}
g =  2 \, \cdot \frac{M (4-\Xi) + 4 Q (1-\Xi)}{2M + Q (2-\Xi)}\,\,
\label{equalgyro}
\end{equation}
It follows that in the limit $ \Xi \rightarrow 0 $, when the rotation occurs at the speed of light, the gyromagnetic ratio $ g= 4 $ independent of the electric charge. This is precisely the same result as that obtained in \cite{aliev4} for small perturbative values of the electric charge. It also confirms our earlier  claim that in the critical limit $ \Xi \rightarrow 0 $, the value of the gyromagnetic ratio obtained for small electric charge will remain  unchanged for generic values of the charge \cite{aliev3}.\\
\noindent {iii)}~ $ l\rightarrow \infty $,~~$ \Xi_a=\Xi_b = 1, $ zero  cosmological constant. In this case equations (\ref{gyroa}) and  (\ref{gyrob}) give the gyromagnetic ratios
\begin{eqnarray}
\label{zerolambda1}
g_{a} & = &  3 \left(1- \frac{Q b}{2 a M + Q b}\right)\,,
\\  [2mm]
g_{b} & =&   3 \left(1- \frac{Q a}{2 b M + Q a}\right)\,,
\label{zerolambda2}
\end{eqnarray}
which for small values of the electric charge, $ Q \rightarrow 0, $ tend to $ g \rightarrow 3 $. This agrees with our previous result obtained in \cite{af} for a weakly charged five-dimensional Myers-Perry black hole. We see that for given other parameters the growth of electric charge acts to decrease the value of the gyromagnetic ratios. Some related numerical analysis as well as the study of the case with different Chern-Simons coupling constant can be found in \cite{kunzfor1, kunzfor2}.

It is also interesting to consider a supersymmetric case in the BPS limit. The BPS limit, as shown in \cite{cclp}, arises when the condition
\begin{eqnarray}
Q & = &  \frac{M}{1+ (a+b) l^{-1}}
\label{bps}
\end{eqnarray}
is fulfilled.  In this limit,  the gyromagnetic ratio in (\ref{equalgyro}) reduces to
\begin{equation}
g =  2 \cdot \frac{\left(4-\Xi\right)\left(1+2\sqrt{1-\Xi}\right) + 4 \left(1-\Xi\right)}{4-\Xi + 4 \sqrt{1-\Xi}}\,\,.
\label{equalgyrobps}
\end{equation}
It follows that for a supersymmetric black hole with equal rotation parameters, the gyromagnetic ratio $ g=4 $  achieved for critical rotation $ \Xi \rightarrow 0 $  $(a \rightarrow l) $ decreases to its value $ g=2 $ for  $\Xi \rightarrow 1 $  $(a \ll l) $.

\section{Conclusion}

In past years the gyromagnetic ratio of stationary and axisymmetric solutions of the Einstein-Maxwell equations  has  been the subject of
many studies in asymptotically flat case. The present paper extends previous results on the spacetimes with asymptotically non-flat structure. We have considered the most simple example of the asymptotically non-flat spacetimes, namely, the Kerr-Newman spacetime with a straight cosmic string passing through its poles. The asymptotic behavior of this spacetime is conical. We have shown that the gyromagnetic ratio still remains $ g=2 $ though the conserved parameters of the spacetime metric change  due to its conical nature. We have also shown that the  Kerr-Newman-Taub-NUT metric, whose NUT charge makes its asymptotic  structure non-flat, has the gyromagnetic ratio $ g=2 $. Finally, we have considered a general class of metrics for which the Hamilton-Jacobi and wave equations admit separation of variable. Within this Carter's separability ansatz we have shown that
all the metrics in this class (with both NUT charge and cosmological constant)  must have the gyromagnetic ratio $ g=2 $. This result fills a long standing gap in the study of the gyromagnetic ratio in general relativity by extending it to include  the NUT charge and the cosmological constant.

In the final section we have examined the gyromagnetic ratio for recently-found general solution in minimal five-dimensional gauged supergravity. For certain  ranges of its physical parameters, this solution represents a rotating charged black hole with two independent angular momenta and asymptotic AdS structure. We have obtained the general expressions for two gyromagnetic ratios of the black hole. Exploring some special cases, we have found that when the black hole with equal angular momenta is rotating at the speed of light, its gyromagnetic ratio becomes $ g=4 $ independent of the electric charge. This is in agreement with our previous result obtained for small perturbative values of the electric charge. We have also found that for a supersymmetric case in the BPS limit the gyromagnetic ratio ranges in the interval $ 2\leq  g \leq4\, .$

\section{Acknowledgments}

The author is grateful to the Scientific and Technological Research Council of Turkey (T{\"U}B\.{I}TAK)  for partial support through the
Research Project 105T437. He is also grateful to Teoman Turgut and Nihat Berker for their continuous stimulating encouragements.

\end{document}